\documentclass[showpacs,preprintnumbers,prd,nofootinbib,floats,amssymb,floatfix]{revtex4}
\usepackage{amsmath}
\usepackage{amsfonts}
\usepackage{graphicx}
\usepackage{hyperref}
\usepackage{amsmath}
\usepackage{amstext}
 
\begin{document}

\title {Symplectic analysis of three dimensional   Abelian topological   gravity }
\author{R. Cartas-Fuentevilla} \email{rcartas@ifuap.buap.mx}
\author{Alberto Escalante}  \email{aescalan@ifuap.buap.mx}
\author{Alfredo Herrera-Aguilar} \email{aherrera@ifuap.buap.mx}
 \affiliation{  Instituto de F{\'i}sica, Benem\'erita Universidad Aut\'onoma de Puebla, \\
 Apartado Postal J-48 72570, Puebla Pue., M\'exico, }
\begin{abstract}
A detailed  Faddeev-Jackiw quantization of  an Abelian topological gravity is performed;  we  show that this formalism  is equivalent  and more  economical  than  Dirac's method. In particular, we identify  the complete set of  constraints of the theory,  from which the number of physical degrees of freedom is explicitly computed. We prove that the generalized Faddeev-Jackiw brackets  and the Dirac  ones coincide to each other. Moreover,  we perform the   Faddeev-Jackiw analysis of   the theory  at the chiral point, and  the full  set  of constraints  and the generalized Faddeev-Jackiw brackets are constructed. Finally we compare  our results with those found in  the literature and we  discuss some remarks and prospects.  
\end{abstract}
 \date{\today}
\pacs{98.80.-k,98.80.Cq}
\preprint{}
\maketitle
\section{Introduction}
It is well-known that the unification between quantum mechanics and gravity   is a difficult task to perform. The two promising approaches for  solving the problems that emerge  in the quantum  formulation  of  gravity, namely,    string theory and  loop quantum gravity are, unless still now, in progress \cite{1, 1a,  2, 2a}. Due to  the difficulties   found in the quantum formulation of gravity, it is convenient try to study      the classical and the quantum formulation of toy models for testing  ideas about  actual  gravity theories.  In this respect, there are a lot of interesting  models that have been useful for this aim, as for instance, three dimensional gravity \cite{3, 4, 5, 6, 7}, topological formulations of gravity in three and four dimensions  and the well-known Abelian gravity theories \cite{8, 9, 10, 11}. With respect to three dimensional gravity, there is a toy model very close   to pure  gravity,  the so-called  topologically massive gravity [TMG]  \cite{15, 16, 16a, 16b, 17, 18, 19}. This model  describes the propagation of a massive graviton and  black hole  solutions obeying the laws of black hole thermodynamics,   is Poincar\'e gauge invariant  and also  provides at the chiral point  a description of the structure of an $AdS_3$ asymptotic spacetime  \cite{20}. However,  in spite of  the works found in the literature on this model,  the classical and quantum analysis  is still a subject of research. In fact,  the classical and quantum study of this model  is also a difficult task to perform; from the point of view of Dirac's approach, the correct identification of the constraints between first and second class is  a difficult work to carry out \cite{21} and we know  the correct identification of the constraints is both  the best guideline to carry out the correct counting of physical  degrees of freedom and the first step to perform the  quantum treatment of the theory. From the quantum point of view TMG with negative cosmological constant is unstable, massive gravitons and black holes have negative energy. An exception is present when the theory is analyzed at the chiral point, where   the black holes and gravitons have non-negative masses and the  linearized gravitational excitations around $AdS_3$ is stable \cite{21}.    In this manner, new developments focussed to the  classical and the quantum analysis of TMG   by using a different framework to the Dirac one  come to be  relevant. In this respect, in the present paper we study an Abelian analog of TMG by using the Faddeev-Jackiw [FJ] method being  a powerful    alternative approach for studying singular systems \cite{15-c, 15c1, 15c2, 15c3, 15c4, 15c5, 15c6, 15c7}. In fact, the  FJ method   is a symplectic approach, namely, all  relevant information of the theory can be obtained through an invertible symplectic tensor, which is  constructed by means  of the symplectic variables that are identified as the degrees of freedom.  Because of the theory  is singular  there will be constraints,  and FJ has the   advantage that  all  constraints are at the same footing, since   it is not necessary  to perform  the  classification of the constraints in primary, secondary, first class or second class  as in Dirac's method is done \cite{22, pons, 23, 24, 25, 26, 27, 28, selfdual, 29, 30, 31}. When  the symplectic tensor is obtained, then  its components are identified with the FJ  generalized  brackets;  Dirac's brackets and FJ brackets coincide to each other. On the other hand, we have chosen  to analyze  the Abelian analog of TMG because  it is well-known  that Abelian models of gravity in four and three dimensions  have interesting features. In fact,  in the  four dimensional case, there exists the so-called $SU(2)$ gravity  (or Husain-Kuchar gravity) \cite{12}; the model itself is closely related to Ashtekar's formulation of general relativity in terms of the new canonical variables. Moreover, $SU(2)$ gravity  has three degrees of freedom per space-time point, and  is devoid  of  the Hamiltonian constraint,   and this fact makes the theory  more easy to quantize where the physical states for general relativity form a subset
of the states of the theory under study. On the other hand,  an alternative description for   Abelian gravity has been  proposed in \cite{13}. In fact, in \cite{13} the model is a $U(1)$ gauge invariant theory, with  zero degrees of freedom and reducibility among the constraints, and  the quantum formulation  shares the same path-integral formula with that of  $BF$ theory.  Furthermore,  an Abelian description of gravity is obtained in the limit  $G\rightarrow0$  of the four dimensional Palatini's theory  \cite{32a, 32}. In fact, the theory presents reducible constraints  and lacks  physical degrees of freedom. With respect  to the three dimensional  case, we find in the literature the so-called 2+1 gravity without dynamics \cite{14},  describing   a generally covariant  topological theory   without  Hamiltonian constraint. In fact, the theory only presents  the spatial diffeomorphism constraint  and the quantum description can be carried  out  in an elegant form by  using the loop representation. Finally,  we find also in the literature  the 2+1 Abelian topological massive gauge theories, where  the so-called SD model and Maxwell Chern-Simons theory  are the subjects of several works (see \cite{33} and cites there in).    \\
In this manner, with the antecedents commented above, in this paper the FJ analysis of an  Abelian version  of TMG is performed. We find  the full set of  constraints and  the generalized FJ brackets are  constructed. Furthermore, we study the theory at the chiral point and by using the symplectic approach  the symmetries of the theory are revealed. Finally,  a pure Dirac's method applied to TMG at the chiral point is  added. We report the complete set of  first and second class constraints, then we construct the fundamental Dirac's brackets and we show that the Dirac and FJ brackets coincide to each other. 

\section{Symplectic formalism of   three dimensional Abelian topological  gravity}
\maketitle
 As it was commented above, the model that we shall  study  in this section is an Abelian version  of TMG.  In spite of  TMG has been analyzed in the context of Dirac's  approach,  in the process one finds   several problems in order to identify  which  constraints  of the theory are  first class or second class  \cite{21}. In this manner, in this section we will study a toy model for testing ideas about classical gravity that could be  useful either in the three dimensional or   four-dimensional gravity theory. Hence, our laboratory is given by an  Abelian version of TMG, and our tool is  the symplectic approach developed by  FJ  for revealing the fundamental symmetries and  the constraints of the theory. We start from  the well-known TMG action  given by  \cite{19}
\begin{eqnarray}
S[A,e,\lambda]&=&\int_{\mathcal{M}}\left[2 e^{i}\wedge F_{i}[A]+\lambda^{i}\wedge T_{i}+\frac{1}{\mu}A^{i}\wedge \left(dA_{i}+\frac{G}{3}f_{ijk}A^{j}\wedge A^{k}\right)\right], 
\label{1}
\end{eqnarray}
where  $A^{i}=A_{\mu}{^{i}}dx^{\mu}$ is a connection 1-form valued on the adjoint representation of the Lie group $SO(2,1)$, which admits an invariant totally anti-symmetric tensor $f_{ijk}$,      $e^{i}=e_{\mu}^{i}dx^{\mu}$ is a triad 1-form that represents the  gravitational field and  $F^{i}$ is the curvature 2-form of the connection $A^{i}$, i.e.,  $F_{i}\equiv dA_{i}+\frac{G}{2}f_{ijk}A^{j}\wedge A^{k}$ and $G$ is the gravitational coupling constant. Finally,  $\lambda^{i}$ are  Lagrange multiplier 1-forms that ensure that  the torsion vanishes $T_{i}\equiv de_{i}+ Gf_{ijk}A^{j}\wedge e^{k}=0$;  $x^{\mu}$ are the coordinates that label the points of the 3-dimensional manifold $\mathcal{M}$. In our notation, Greek letters are indices for the spacetime and run from 0 to 2, and $a, b, c=1, 2$ are space indices, the middle latin alphabet letters  $(i,j,k,...)$   are associated with  the internal group $SO(2, 1)$ and run from $1$ to $3$.\\
 The gravitational coupling constant    has been  introduced in oder to take the $G \rightarrow0$ limit, and to obtain  an  Abelian version of TMG, something similar  can be found in \cite{32} where  the FJ analysis of an Abelian analog of four-dimensional Palatini's theory was reported. By taking the $G \rightarrow 0$ limit we obtain the following action 
\begin{eqnarray}
S[A,e,\lambda]&=&\int_{\mathcal{M}}\left[2 e^{i}\wedge F[A]_{i}+\lambda^{i}\wedge T_{i}+\frac{1}{\mu}A^{i}\wedge dA_{i}\right], 
\label{1s}
\end{eqnarray}
where $F{_{ab}}^{i}=\partial_{a}A^{i}{_{b}} - \partial_{b}A^{i}{_{a}}$,  $T^{i}{_{ab}}=\partial_{a}e^{i}{_{b}} - \partial_{b}e^{i}{_{a}}$ and now  the dynamical variables  are a collection of $U(1)$ gauge invariant fields. \\
The equations of motion obtained from the action (\ref{1s}) are given by 
\begin{eqnarray}
\epsilon^{\alpha\nu\rho}\left(2 F_{\nu\rho}{^{i}}+\partial_{\nu}\lambda_{\rho}{^{i}}\right)&=&0,\nonumber\\
\epsilon^{\alpha\nu\rho}\left(2 T_{\nu\rho}{^{i}}+\epsilon^{i}{_{jk}}\lambda_{\nu}{^{j}}e_{\rho}{^{k}}+2\mu^{-1}F_{\nu\rho}{^{i}}\right)&=&0,\nonumber\\
\epsilon^{\alpha\nu\rho}T_{\nu\rho}{^{i}}&=&0,
\label{eqm}
\end{eqnarray}
we can see that the equation of motion related with the torsion  $\partial_{\alpha}e^{i}{_{\beta}} - \partial_{\beta}e^{i}{_{\alpha}}=0$,  implies that $e^i_\alpha = \partial_\alpha f^i $, thus,  the  background scenario corresponds locally to Minkowski spacetime, the model  shares similarities with the Abelian version of Palatini's theory reported in \cite{32a}. \\
On the other hand,  by performing the $2+1$ decomposition of the fields, the  action takes the following form 
\begin{eqnarray}
{\mathcal{L}}^{(0)}&=& \int\ \Big(2 \varepsilon^{ab}e^{i}{_{b}}\dot{A}{_{ai}} + \frac{1}{\mu}\varepsilon^{ab}A^{i}{_{b}}\dot{A}_{ai} + \varepsilon^{ab}\lambda_{ib}\dot{e}^{i}{_{a}}  - V^{(0)} \Big) dx^{3}, 
\label{eq2}
\end{eqnarray}
where $\varepsilon^{0ab}\equiv \varepsilon^{ab}$ is the antisymmetric tensor and    $V^{(0)}=-e^{i}{_{0}} \left[ F_{abi}\varepsilon^{ab} + \varepsilon^{ab}\partial_{a}\lambda_{bi} \right] - A^{i}{_{0}}\left[ \varepsilon^{ab}T_{abi} + \frac{1}{\mu}\xi^{ab}F_{abi} \right] -\frac{\lambda^{i}{_{0}}}{2}\left[\varepsilon^{ab}T_{abi}\right]$ is  identified as  the symplectic potential (see the appendix B). Hence, in the following lines  we will study the action (\ref{eq2})  within the context of Faddeev-Jackiw.  In order to perform this aim, from (\ref{eq2}) we identify the following symplectic variables   $\overset{(0)}{\xi}{^{A}}= \left(A^{i}{_{a}}, A^{i}{_{0}}, e^{i}{_{a}}, e^{i}{_{0}}, \lambda^{i}{_{a}}, \lambda^{i}{_{0}} \right)$ \, and the 1-form $\overset{(0)}{a}{_{B}}= \left(2\varepsilon^{ab}e{_{ib}} + \frac{1}{\mu}\varepsilon^{ab}A_{ib}, 0, \varepsilon^{ab}\lambda_{ib}, 0, 0, 0 \right)$, here $A, B, C= 1, 2, 3..., $ label the number of field variables,  for instance, there are 27 field variables and all  them are represented by $\overset{(0)}{\xi}{^{A}}$.   Thus,  by taking into account these symplectic variables the equations of motion of the action (\ref{eq2}) are given by 
\begin{equation}
f^{(0)}_{A B}\dot{\xi}^{B}=\frac{\partial V^{(0)}(\xi)}{\partial\xi^{A}},
\label{eq36}
\end{equation}
where the symplectic matrix $f^{(0)}_{AB}$ takes the form
\begin{equation}
f^{(0)}_{AB}(x,y)=\frac{\delta a_{B}(y)}{\delta\xi^{A}(x)}-\frac{\delta a_{A}(x)}{\delta\xi^{B}(y)},
\label{37}
\end{equation}
and it is given explicitly  by 
\begin{eqnarray}
\label{eq}
f^{(0)}_{AB}=
\left(
  \begin{array}{cccccc}
 \frac{2}{\mu}\varepsilon^{ab}\eta_{ij}    &	\quad   0   &\quad  -2\varepsilon^{ab}\eta_{ij}   &\quad  0     & \quad 0  &\quad  0 	 	 \\                                                                        
 0 &\quad   0   &\quad  0   &\quad   0 &\quad   0   &\quad   0 \\                                                                   
2\varepsilon^{ab}\eta_{ij}   &\quad  0   &\quad   0      &\quad   0   &\quad    -\varepsilon^{ab}\eta_{ij}   &\quad  0 	\\
    0   &\quad  0   &\quad   0    &\quad 0  &\quad  0	 &\quad  0 \\
0  &\quad  0  &\quad    \varepsilon^{ab}\eta_{ij}  &\quad  0  &\quad  0 	&\quad  0 	 \\
0   &\quad   0  &\quad  0  &\quad  0   &\quad  0 	&\quad  0 \\
 \end{array}
\right) \delta^{2}(x-y), 
\end{eqnarray}
we can observe that this matrix is singular and therefore  there will  constraints. Because of the symplectic matrix is singular, it has the following null-vectors  $\mathcal{V}^{}{_{1}}= \left (0, v^{A^{i}{_{0}}}, 0, 0, 0, 0 \right), \quad \mathcal{V}^{}{_{2}}= \left(0, 0, 0, v^{e^{i}{_{0}}}, 0, 0 \right) $ \, and \, $ \mathcal{V}^{}{_{3}} = \left(0, 0, 0, 0, 0, v^{\lambda^{i}{_{0}}} \right)$, where $v^{A^{i}}, v^{e^{i}{_{0}}}, v^{\lambda^{i}{_{0}}}  $ are arbitrary functions. From those modes, we obtain the following constraints \cite{23}
\begin{eqnarray}
\Omega^{(0)}_{i}&=& \int\ dx^{2}\mathcal{V}^{A}{_{1}}\frac{\delta}{\delta\xi^{A}} \int\ dy^{2}V^{(0)}(\xi) =\varepsilon^{ab}T_{abi} + \frac{1}{\mu}\varepsilon^{ab}F_{abi}=0, \\ 
\beta^{(0)}_{i}&=& \int\ dx^{2}\mathcal{V}^{A}{_{2}}\frac{\delta}{\delta \xi^{A}} \int\ dy^{2}V^{(0)}(\xi)= \varepsilon^{ab}F_{abi} + \varepsilon^{ab}\partial_{a}\lambda_{bi}=0, \\ 
\Sigma^{(0)}_{i}&=& \int\ dx^{2}\mathcal{V}^{A}{_{3}}\frac{\delta}{\delta \xi^{A}} \int\ dy^{2}V^{(0)}(\xi)=\varepsilon^{ab}T_{abi}=0.
\end{eqnarray}
In order to observe if there are more constraints, we construct the following matrix (see Appendix B)
\begin{eqnarray}
\bar{f}_{AB}\dot{\xi}^{A}&=& Z_{B}(\xi), 
\label{eq9}
\end{eqnarray}
where

\begin{eqnarray}
\label{eq}
 \bar{f}_{AB}=
\left(
\begin{array}{cc} 
f^{(0)}_{AB} \\ 
\frac{\delta\Omega^{(0)}}{\delta\xi^{A}} 
\end{array}
\right),
\label{eq10}
\end{eqnarray} 

and

\begin{eqnarray}
\label{eq}
 Z_{A}(\xi)=
\left(
\begin{array}{ccc} 
\frac{\partial V^{(0)}}{\partial \xi^A} \\ 
0 \\
0\\ 
\end{array}
\right), 
\label{eq11}
\end{eqnarray}
hence $\bar{f}_{AB}$ is given by
\begin{eqnarray}
\label{eq}
\bar{f}_{AB}=
\left(
  \begin{array}{cccccc}
 \frac{2}{\mu}\varepsilon^{ab}\eta_{ij}    &	\quad   0   &\quad  -2\varepsilon^{ab}\eta_{ij}   &\quad  0     & \quad 0  &\quad  0 	 	 \\                                                                        
 0 &\quad   0   &\quad  0   &\quad   0 &\quad   0   &\quad   0 \\                                                                   
2\varepsilon^{ab}\eta_{ij}   &\quad  0   &\quad   0      &\quad   0   &\quad    -\varepsilon^{ab}\eta_{ij}   &\quad  0 	\\
    0   &\quad  0   &\quad   0    &\quad 0  &\quad  0	 &\quad  0 \\
0  &\quad  0  &\quad    \varepsilon^{ab}\eta_{ij}  &\quad  0  &\quad  0 	&\quad  0 	 \\
0   &\quad   0  &\quad  0  &\quad  0   &\quad  0 	&\quad  0 \\
\frac{2}{\mu}\eta_{ij}\varepsilon^{ab}\partial_{a}   &\quad  0  &\quad  2\eta_{ij}\varepsilon^{ab}\partial_{a}  &\quad 0 &\quad 0 &\quad 0  \\
2\varepsilon^{ab}\partial_{a}\delta_{ij}  &\quad 0 &\quad 0 &\quad 0 &\quad \eta_{ij}\varepsilon^{ab}\partial_{a} &\quad 0 \\
0 &\quad 0 &\quad \delta_{ij}2\varepsilon^{ab}\partial_{a} &\quad 0 &\quad 0 &\quad 0 \\
 \end{array}
\right) \delta^{2}(x-y),
\label{eq12}
\end{eqnarray} 
the matrix (\ref{eq12}) is not a square matrix and   still has null vectors. The modes of the matrix (\ref{eq12}) are given by
\begin{eqnarray}
\bar{\mathcal{V}}^{A}{_{1}}&=& \left(\partial_{a}V^{\lambda^{i}}, V^{\lambda ^{ A^{i}{_{0}}}}, 0 , 0, 0, 0, V^{\lambda^{i}}, 0, 0 \right), \\
\bar{\mathcal{V}}^{A}{_{2}}&=& \left(0, 0, 0, V^{\lambda^{ e^{i}{_{0}}}}, 2\partial_{a}V^{\lambda^{i}}, 0, 0, 0, -V^{\lambda}\right), \\
\bar{\mathcal{V}}^{A}{_{3}}&=& \left(0, 0, \partial_{a}V^{\lambda^{i}}, 0, 0, V^{\lambda^{i}{_{0}}}, 0, V^{\lambda}, 0 \right).
\end{eqnarray} 
On the other hand,  $Z_{A}$ takes the form
\begin{eqnarray*}
\label{eq}
Z_{A}=
\left(
  \begin{array}{c}
 \varepsilon^{ab}\partial_{a}e_{0i} + \frac{1}{\mu}\varepsilon^{ab}\partial_{a}A_{0i} 
\\ - \left[\varepsilon^{ab}T_{abi} + \frac{1}{\mu}\varepsilon^{ab}F_{abi}\right] \\ 
\varepsilon^{ab}\partial_{a}A_{0i} + \varepsilon^{ab}\partial_{a}\lambda_{0i} 
 \\- \left[ F_{abi}\varepsilon^{ab} + \varepsilon^{ab}\partial_{a}\lambda_{bi}\right] \\
\varepsilon^{ab}\partial_{a}e_{0i} \\ 
-\frac{\varepsilon^{ab}}{2}T_{abi} \\
0 \\
0 \\
0 \\
 \end{array}
\right).
\end{eqnarray*} \\
The contraction of the null vectors with $Z_{A}$, namely, $\bar{\mathcal{V}}^{A}Z_{A}|_{\Omega_i ^{(0)}, \beta_i^{(0)}, \Sigma_i ^{(0)}=0} = 0$ yields   identities. Hence, there are not more constraints in the theory (see the  Appendix B). 
With  this   information, we construct a new symplectic Lagrangian given by
\begin{eqnarray}
{\mathcal{L}}{^{(1)}} &=& [2\varepsilon^{ab}e_{bi} + \frac{1}{\mu}\varepsilon^{ab}A_{bi}]\dot{A}^{i}{_{a}} +\varepsilon^{ab}\lambda_{bi}\dot{e}^{i}{_{a}} - [ \varepsilon^{ab}F_{abi} + \varepsilon^{ab}\partial_{a}\lambda_{bi}]\dot{\alpha}^{i} \nonumber \\
& -& [\varepsilon^{ab}T_{abi} + \frac{1}{\mu}\varepsilon^{ab}F_{abi}]\dot{\beta}{^{i}} -[\varepsilon^{ab}T_{abi}]\dot{\Gamma}^{i} - V^{(1)}, 
\label{eq16}
\end{eqnarray}
where we have taken  \, $e^{i}{_{0}}=\dot{\alpha}^{i},\, A^{i}{_{0}}=\dot{\beta}^{i},\, \lambda^{i}{_{0}}=\dot{\Gamma}^{i}$  as  a set of  Lagrange multipliers enforcing the constraints with a vanishing potential  $V^{(1)}=V^{(0)}\mid_{\Omega^{(0)}_{i}=0,\beta^{(0)}_{i}=0, \Sigma^{(0)}=0}=0$, which  reflects the general covariance of the theory. In this manner,  from (\ref{eq16}) we identify the following symplectic variables 
$\overset{(1)}{\xi}{^{A}}= \left(A^{i}{_{a}}, \beta^{i}, e^{i}{_{a}}, \alpha^{i}, \lambda^{i}{_{a}}, \Gamma^{i}\right)$\, and the 1-forms $\overset{(1)}{a}{^{A}}= \left(2\varepsilon^{ab}e_{bi} + \frac{1}{\mu}\varepsilon^{ab}A_{bi}, -\varepsilon^{ab}T_{abi} - \frac{1}{\mu}\varepsilon^{ab}F_{abi}, \varepsilon^{ab}\lambda_{bi}, -\theta\varepsilon^{ab}F_{abi} - \varepsilon^{ab}\partial_{a}\lambda_{bi}, 0, \varepsilon^{ab}T_{abi}\right)$. By using these symplectic variables, we find the following symplectic matrix
\begin{eqnarray}
\label{eq}
f^{(1)}_{AB}=
\left(
  \begin{array}{cccccc}
 \frac{2}{\mu}\varepsilon^{ab}\eta_{ij}    &	\quad   -\frac{2}{\mu}\eta_{ij}\varepsilon^{ab}\partial_{a}  &\quad  -2\varepsilon^{ab}\eta_{ij}   &\quad  -2\eta_{ij}\varepsilon^{ab}\partial_{a}     & \quad 0  &\quad  0 	 	 \\                                                                        
\frac{2}{\mu} \eta_{ij} \varepsilon^{ab}\partial_{a} &\quad   0   &\quad  -2\eta_{ij}\varepsilon^{ab}\partial_{a}   &\quad   0 &\quad   0   &\quad   0 \\                                                                   
-2\varepsilon^{ab}\eta_{ij}   &\quad  -2\eta_{ij}\varepsilon^{ab}\partial_{a}   &\quad   0      &\quad   0   &\quad    \varepsilon^{ab}\eta_{ij}   &\quad  2\eta_{ij}\varepsilon^{ab}\partial_{a} 	\\
 2\eta_{ij}\varepsilon^{ab}\partial_{a}   &\quad  0   &\quad   0    &\quad 0  &\quad  \eta_{ij}\varepsilon^{ab}\partial_{a}	 &\quad  0 \\
0  &\quad  0  &\quad    -\varepsilon^{ab}\eta_{ij}  &\quad  -\eta^{ij}\varepsilon^{ab}\partial_{a}  &\quad  0 	&\quad  0 	 \\
0   &\quad   0  &\quad  -2\eta_{ij}\varepsilon^{ab}\partial_{a}  &\quad  0   &\quad  0 	&\quad  0 \\
 \end{array}
\right) \nonumber \\
\times \delta^2(x-y).
\label{mat19}
\end{eqnarray} 
Where we can observe  that this matrix is singular. However, we have shown  that there  are not more constraints,  hence  the system has a gauge symmetry. In fact,  the gauge symmetry  is encoded in the null vectors of the matrix (\ref{mat19}). It is straightforward to see  that  the following Abelian gauge symmetries  are  obtained from    the null vectors of the above matrix
\begin{eqnarray}
A_{\mu}^i\rightarrow A^i_{\mu} + \partial_\mu \kappa^i \nonumber \\
e_\mu^i\rightarrow e_\mu^i+ \partial_\mu \tilde{\kappa}^i, 
\end{eqnarray}
where $\kappa^i$ and $\tilde{\kappa} ^i$ are gauge parameters. Hence, in order to obtain a symplectic tensor, we will fix the gauge, and we choose the temporal gauge
\begin{eqnarray*}
A^{i}{_{0}}&=& 0, \\
e^{i}{_{0}}&=& 0, \\
\lambda^{i}{_{0}}&=& 0, 
\end{eqnarray*} 
this means that $\beta^{i}=cte,\, \alpha^{i}=cte$ \, and \, $\Gamma^{i}=cte$.  In this manner, with  that information  we construct the following  new symplectic Lagrangian
\begin{eqnarray}
{\mathcal{L}}^{(2)} &=& [2\varepsilon^{ab}e_{bi} + \frac{1}{\mu}\varepsilon^{ab}A_{bi}]\dot{A}^{i}{_{a}} + \varepsilon^{ab}\lambda_{bi}\dot{e}^{i}{_{a}} - [ F_{abi}\varepsilon^{ab} + \varepsilon^{ab}\partial_{a}\lambda_{bi}]\dot{\alpha}^{i} + \alpha^{i}\dot{\rho}_{i} \nonumber \\
&-& [\varepsilon^{ab}T_{abi} + \frac{1}{\mu}\varepsilon^{ab}F_{abi}]\dot{\beta}^{i} - \beta^{i}\dot{\varpi}_{i} - [\varepsilon^{ab}T_{abi}]\dot{\Gamma}^{i} - \Gamma^{i}\dot{r}^{i}, 
\label{eq18}
\end{eqnarray}
where we have introduced new Lagrange multipliers enforcing the gauge fixing, namely, $\varpi_i, \Gamma^i, r^i$. Hence, from (\ref{eq18}) we identify the following symplectic variables\, $\overset{(2)} {\xi^{A}}= \left(A^{i}{_{a}}, \beta^{i}, e^{i}{_{a}}, \alpha^{i}, \lambda^{i}{_{a}}, \Gamma^{i}, \rho_{i}, \omega_{i}, r_{i}\right)$ \, and the 1-forms  $\overset{(2)}{a}{^{B}} =(2\varepsilon^{ab}e_{bi} + \frac{1}{\mu}\varepsilon^{ab}A_{bi}, 
- \varepsilon^{ab}T_{abi} - \frac{1}{\mu}\varepsilon^{ab}F_{abi} + \varpi_{i}, \varepsilon^{ab}\lambda_{bi} - \varepsilon^{ab}F_{abi} - \varepsilon^{ab}\partial_{a}\lambda_{bi} + \rho_{i}, 0, -\varepsilon^{ab}T_{abi} + r_{i}, 0, 0, 0 ,0)$. 
In this manner, the symplectic matrix reads 

\begin{eqnarray}
\label{eq}
\overset{(2)}{f}_{AB}=
\left(
  \begin{array}{ccccccccc}
 \frac{2}{\mu}\varepsilon^{ab}\eta_{ij}    	&\,   -\frac{2}{\mu}\delta_{ij}\varepsilon^{ab}\partial_{a} 	&\,	 -2\varepsilon^{ab}\eta_{ij}       &\,  -2\eta_{ij}\varepsilon^{ab}\partial_{a}     & \, 	0  	&\,	  0  	&\,	 0	  &\,	 0	  &\,	 0	 	 \\                                                                        
 \frac{2}{\mu}\eta_{ij}\varepsilon^{ab}\partial_{a}	&\,	   0	   &\,  -2\eta_{ij}\varepsilon^{ab} 	  &\,  	 0 	&\, 	0   	&\,  	 0 	 &\,   0 	&\,  	-\eta^{i}{_{j}}	  &\, 	 0	\\                                                                   
2\eta_{ij}\varepsilon^{ab}  	 &\,	  2\eta_{ij}\varepsilon^{ab}  	 &\, 	  0      &\,	   0   &\,  	  -\varepsilon^{ab}\eta_{ij}  	 &\,  -2\varepsilon^{ab}\partial_{a}	&\, 	 0	&\,	0	&\,	0	\\
 2\eta_{ij}\varepsilon^{ab}\partial_{a}	   &\, 	 0 	  &\, 	  0  	  &\,	 0	  &\,  \eta_{ij}\varepsilon^{ab}\partial_{a}		 &\,  	0 	&\,	-\eta^{i}{_{j}}	&\,	0	&\,	0	\\
0  	&\,  	0  	&\,    \varepsilon^{ab}\eta_{ij}  	&\,	  -\eta_{ij}\varepsilon^{ab}\partial_{a} 	 &\,	  0 	&\,  	0 	&\, 	0	&\, 	0	&\, 	0		 \\
0   	&\,  	 0	  &\,  	2\varepsilon^{ab}\partial_{a} 	 &\, 	0	   &\,  	0 	&\, 	 0 	&\,	0	&\,	0	&\,	-\eta^{i}{_{j}}	 \\
0	&\,	0	&\,	0	&\,	\eta^{i}{_{j}}	&\,	0	&\,	0	&\,	0	&\,	0	&\,	0	\\
0	&\,	\eta^{i}{_{j}}	&\,	0	&\,	0	&\,	0	&\,	0	&\,	0	&\,	0	&\,	0	\\
0	&\,	0	&\,	0	&\,	0	&\,	0	&\,	\eta^{i}{_{j}}	&\,	0	&\,	0	&\,	0	\\
 \end{array}
\right)  \nonumber \\
\times  \delta^{2}(x-y).  \nonumber \\
\end{eqnarray} 

We can observe that this matrix is not singular, hence, it is a symplectic tensor. The inverse of $\overset{(2)}f_{AB}$ is given by

\begin{eqnarray}
\label{eq}
\left(\overset{(2)}{f}_{AB}\right )^{-1}=
\left(
  \begin{array}{ccccccccc}
 \frac{\mu}{2}\varepsilon_{ab}\eta^{ij}    	&\,  0 	&\,	0      &\,  0     & \, 	-\mu\eta^{i}{_{j}}\varepsilon_{ab} 	&\,	  0  	&\,	 0	  &\,	 \eta^{i}{_{j}}\partial_{a}	  &\,	 0	 	 \\                                                                        
 0	&\,	   0	   &\,  0 	  &\,  	 0 	&\, 	0   	&\,  	 0 	 &\,  		0	&\,  	\eta^{i}{_{j}}	  &\, 	 0	\\                                                                   
0 	 &\,	0  	 &\, 	  0      &\,	   0   &\,  	  \varepsilon_{ab}\eta^{i}{_{j}} 	 &\,  		0	 &\, 	\eta^{i}{_{j}}\partial_{a}	&\,	0	&\,	0	\\
 0	   &\, 	 	0 	  &\, 	  0  	  &\,	 0	  &\,  		0		&\,  		0 		&\,		\eta^{i}{_{j}}	&\,	0	&\,	0	\\
\mu \eta^{i}{_{j}}\varepsilon_{ab}	  	&\,  	0  	&\,    	-\varepsilon_{ab}\eta^{i}{_{j}} 	&\,		  0	 &\,	  2\mu\varepsilon_{ab}\eta^{ij}	&\,  	0 	&\, 	0	&\, 	2\eta^{i}{_{j}}\partial_{a}	&\, 	-2\eta^{i}{_{j}}\partial_{a}		 \\
0   	&\,  	 0	  &\,		0 	 &\, 	0	   &\,  	0 	&\, 	 0 	&\,	0	&\,	0	&\,	\eta^{i}{_{j}}	 \\
0	&\,	0	&\,	\eta^{i}{_{j}}\partial_{a}	&\,	-\eta^{i}{_{j}}	&\,	0	&\,	0	&\,	0	&\,	2\varepsilon^{ab}\partial_{a}	&\,	0	\\
-\eta^{i}{_{j}}\partial_{a}		&\,	-\eta^{i}{_{j}}	&\,	0	&\,	0	&\,	-2\eta^{i}{_{j}}\partial_{a}	&\,	0	&\,	2\eta^{i}{_{j}}\varepsilon^{ab}\partial_{a}	&\,	0	&\,	0	\\
0	&\,	0	&\,	0	&\,	0	&\,	2\eta^{i}{_{j}}\partial_{a}	&\,	-\eta^{i}{_{j}}	&\,	0	&\,	0	&\,	0	\\
 \end{array}
\right)  \nonumber \\
\times \delta^2(x-y). \nonumber \\
\end{eqnarray} 
Where we can identify  the following generalized FJ brackets
\begin{eqnarray}
\{\overset{(2)}{\xi^A}{(x)},\overset{(2)}{\xi^B}(y)\}_{FD}=[f^{(2)}_{AB}(x,y)]^{-1},
\end{eqnarray}
thus,  we obtain 
\begin{align}
\{e^{i}{_{a}},e^{j}{_{b}}\}_{FJ}&=0, \nonumber \\
\{A^{i}{_{a}}, A^{j}{_{b}}\}_{FJ}&=\frac{\mu}{2}\varepsilon_{ab}\eta^{ij}\delta^{2}(x-y), \nonumber \\
\{A^{i}{_{a}},\lambda^{j}{_{b}}\}_{FJ}&=-\mu\eta^{ij}\varepsilon_{ab}\delta^{2}(x-y),\nonumber \\
\{\lambda^{i}{_{a}},\lambda^{j}{_{b}}\}_{FJ}&=2\mu\varepsilon_{ab}\eta^{ij}\delta^{2}(x-y),\nonumber \\
\{e^{i}{_{a}},\lambda^{j}{_{b}}\}_{FJ}&=\varepsilon_{ab}\eta^{ij}\delta^{2}(x-y), \nonumber \\
\{A^{i}{_{a}},e^{j}{_{b}}\}_{FJ}&=0.
\end{align}
It is important to remark  that the algebraic structure  of these  brackets coincide with those reported in \cite{19},  where the non-Abelian case was studied. Furthermore, the constraints are not reducible,  which  makes a difference with other models reported in the literature \cite{13}.  Moreover, the theory under study is  a topological one. In fact,  because in the  FJ framework there is no  difference between the constraints, namely, there  does not  exist a  classification of the constraints in first class and second class as in Dirac's method,  in  the FJ framework   the counting of physical degrees of freedom is performed in the usual way; DF= dynamical variables - FJ constraints.  Thus,  for the theory under study there are 18 canonical variables given by $(A^{i}{_{a}}, e^{i}{_{a}},   \lambda_{ia})$  and the following 18 FJ constraints $(\Omega^{(0)}_{i}, \beta^{(0)}_{i}, \Sigma^{(0)}_{i},  A^{i}{_{0}},  e^{i}{_{0}}, \lambda_{i0}  )$, thus, the theory is devoid of physical degrees of freedom. \\
\section{Symplectic formalism for  Abelian topological  gravity at the chiral point}
As it has been commented above, at the chiral point TMG is a theory with interesting properties:  In fact,  either black holes  or gravitons have non-negative masses and the theory is  dual to a holomorphic boundary CFT. The analysis of TMG at the chiral point, has been reported in \cite{20} where   a linearized perturbation around an $AdS_3$ background has been used. Moreover, in \cite{17}  the theory    beyond the linearized  approximation was studied   and a non-perturbative canonical analysis was  performed. With the same spirit, in this work  we will not use a perturbative analysis.  We will show that the FJ  analysis is equivalent and more economical than the Dirac one. At the end of the paper  an  Appendix developing a detailed Dirac's analysis of Abelian TMG at the chiral point has been added, and  we show that the Dirac and the generalized FJ brackets  coincide to each other. \\
 The action describing TMG  can be  written in an alternative way  \cite{17}
\begin{eqnarray}
I[e,\omega, \lambda] &=& \int \left[2e^{i}\wedge F_{i}[\omega]+ \frac{1}{3 l^{2}}\epsilon_{ijk}e^{i}\wedge e^{j}\wedge e^{k} - \frac{1}{\mu}[\omega^{i}\wedge d\omega_{i} + \frac{1}{3}\epsilon_{ijk}\omega^{i}\wedge \omega^{j}\wedge \omega^{k}] + (\lambda^{i} - \frac{e^{i}}{\mu l^{2}})\wedge T_{i}\right],\nonumber \\
\end{eqnarray}
where $F_i[A]= dA_{i}+\frac{1}{2}f_{ijk}A^{j}\wedge A^{k}$   and $T_{i}\equiv de_{i}+ f_{ijk}A^{j}\wedge e^{k}$.  Furthermore,  it is well-known  that under the following redefinition
\begin{align}
A^{i} =\omega^{i} + \frac{e^{i}}{l} ,& &
 \tilde{A}^{i} = \omega^{i} - \frac{e^{i}}{l},
\end{align}
the action can be written in the following way  \cite{17}
\begin{eqnarray}
I[A, \tilde{A}, \lambda] &=& \left(1- \frac{1}{\mu l}\right)\int \left[A^{i}\wedge dA_{i} + \frac{1}{3}\epsilon_{ijk}A^{i}\wedge A^{j}\wedge A^{k}\right] \nonumber \\
&+& \int 2\lambda^{i}\wedge F_{i}[A] - \left(1+ \frac{1}{\mu l}\right)\int \left[\tilde{A}^{i}\wedge d\tilde{A}_{i} + \frac{1}{3}\epsilon_{ijk}\tilde{A}^{i}\wedge \tilde{A}^{j}\wedge \tilde{A}^{k}\right] \nonumber \\
&-& \int 2\lambda^{i}\wedge F_{i}[\tilde{A}].
\end{eqnarray}

Now at the chiral point we take $\mu^{2}l^{2} = 1$ \cite{17, 20}, thus  the Chern-Simons term related  with the dynamical field $A$  is removed and  the action is simplified to
\begin{eqnarray}
I[A,\tilde{A},\lambda]&=& \frac{l}{2}\int 2\lambda^{i}\wedge F_{i}[A] - l\int \left[\tilde{A}^{i}\wedge d\tilde{A}_{i} + \frac{1}{3}\epsilon_{ijk}\tilde{A}^{i}\wedge \tilde{A}^{j}\wedge \tilde{A}^{k}\right] - \frac{l}{2}\int 2\lambda^{i}\wedge F_{i}[\tilde{A}], 
\label{eq32s}
\end{eqnarray}
where  the $\lambda$ field has taken the role of the tetrad-like field coupling  two Einstein-Hilbert copies  depending on  the connections  $A$ and $\tilde A$ or also it can be viewed as two $BF$-like  copies,  where the field $\lambda$ take the roll of the $B$ field. The action (\ref{eq32s}) has been analyzed in \cite{17} by using the  Dirac method  and it has been shown the the action describes the propagation of a physical bulk degree of freedom corresponding,  at the linearized level,  to the topologically   massive graviton. Hence, in this section we will analyze the Abelian analog of the action (\ref{eq32s}) from the symplectic  perspective  in order to obtain in an  easy way   its symmetries.   \\
The Abelian analog of action (\ref{eq32s}) is given by 
\begin{eqnarray}
I[A,\tilde{A},\lambda]&=& \frac{l}{2}\int 2\lambda^{i}\wedge F_{i}[A] - l\int \left[\tilde{A}^{i}\wedge d\tilde{A}_{i} \right] - \frac{l}{2}\int 2\lambda^{i}\wedge F_{i}[\tilde{A}], 
\label{eq32}
\end{eqnarray}
thus, by performing  the 2+1 decomposition we obtain the following Lagrangian density 
\begin{eqnarray}
{\mathcal{L}}^{(0)} &=& l\lambda_{ib}\dot{A}^{i}{_{a}}\epsilon^{0ab} - l(\lambda_{ib}\epsilon^{0ab} + \tilde{A}_{ib}\epsilon^{0ab})\dot{\tilde{A}}^{i}{_{a}} - V^{(0)},
\label{eq33}
\end{eqnarray}
where $F^{i}{_{ab}} $ is defined as above $F^{i}{_{ab}} = \partial_{a}A^{i}{_{b}} - \partial_{b}A^{i}{_{a}}$,  $\tilde{F}^{i}{_{ab}} = \partial_{a}\tilde{A}^{i}{_{b}} - \partial_{b}\tilde{A}^{i}{_{a}}$ and  $V^{(0)}= -lA^{i}{_{0}}\partial_{a}\lambda_{ib}\epsilon^{0ab}  + l\tilde{A}^{i}{_{0}}[\tilde{F}_{iab}\epsilon^{0ab} + \partial_{a}\lambda_{ib}\epsilon^{0ab}] -\frac{1}{2}\lambda_{i0}[F^{i}{_{ab}} - \tilde{F}^{i}{_{ab}}]\epsilon^{0ab}$ is the symplectic potential of the theory. 
Hence,  from (\ref{eq33})  we identify the following symplectic variables $\overset{(0)}{\xi^{A}} = (A^{i}{_{a}}, A^{i}{_{0}}, \tilde{A}^{i}{_{a}}, \tilde{A}^{i}{_{0}}, \lambda_{ia}, \lambda_{i0})$ and the 1-forms $\overset{(0)}{a_{B}} = (l\lambda_{jb}\epsilon^{0ab}, 0, -l(\lambda_{ib}\epsilon^{0ab} + \tilde{A}_{ib}\epsilon^{0ab}),0,0,0)$. By using these symplectic variables,  the  symplectic matrix has the form 
\begin{eqnarray*}
\label{eq34}
\overset{(0)}{f}_{AB}=
\left(
  \begin{array}{cccccc}
0   	&\quad 	   0 	  &\quad	0     	 &\quad  	0     	& \quad 	-l\epsilon^{0ab}\delta^{i}{_{j}} 	&\quad	  0  	 \\                                                                       
0	&\quad	   0	   &\quad    0 	  &\quad  	 0 	&\quad 	0   	&\quad  	 0 	\\                                                                   
0 	 &\quad	   0  	 &\quad 	  -2l\epsilon^{0ab}\eta_{ij}     &\quad	   0   &\quad	 l\epsilon^{0ab}\delta^{i}{_{j}}	 &\quad  		0		\\
0	   &\quad	 	0 	  &\quad 	  0  	  &\quad	 0	  &\quad  		0		&\quad  		0 	\\
l\epsilon^{0ab}\delta^{i}{_{j}}	  	&\quad  	0  	&\quad    	-l\epsilon^{0ab}\delta^{i}{_{j}} 	&\quad		  0	 &\quad	0	&\quad 	0 		 \\
0   	&\quad  	 0	  &\quad		0 	 &\quad 	0	   &\quad 	0 	&\quad 	 0 	 \\
 \end{array}
\right) \delta^2(x-y).
\end{eqnarray*} 
This matrix is singular and has the following modes  $ \mathcal{V}^{(0)}_{1} = (0,v^{A^{i}{_{0}}},0,0,0,0),$ \\ 
$\mathcal{V}^{(0)}_{2} = (0,0,0, v^{\tilde{A}^{i}{_{0}}},0,0),$ \, $\mathcal{V}^{(0)}_{3} = (0,0,0,0,0,v^{\lambda_{i0}})$; by using these  null vectors, we can identify  the following constraints
\begin{eqnarray*}
\Omega^{(0)}_{1} &=& \int dx^{2} \left(\mathcal{V}^{(0)}_{1}\right)^A\frac{\delta}{\delta \overset{(0)} {{\xi}^{A}} } \int d^{2}y V^{(0)}(\xi) \,=\,  l\partial_{a}\lambda_{ib}\epsilon^{0ab} \,= \, 0, \\
\Omega^{(0)}_{2} &=& \int dx^{2} \left(\mathcal{V}^{(0)}_{2}\right)^A\frac{\delta}{\delta\overset{(0)} {{\xi}^{A}} }  \int d^{2}y V^{(0)}(\xi) \,=\, l\{ \epsilon^{0ab}\tilde{F}_{iab} + \partial_{a}\lambda_{ib}\epsilon^{0ab}\} \,=\, 0, \\
\Omega^{(0)}_{3} &=& \int dx^{2}\left(\mathcal{V}^{(0)}_{3}\right)^A \frac{\delta}{\delta\overset{(0)} {{\xi}^{A}} }  \int d^{2}y V^{(0)}(\xi) \,=\, \frac{\epsilon^{0ab}}{2}[F^{i}{_{ab}} - \tilde{F}^{i}{_{ab}}] \,=\, 0.
\end{eqnarray*}
Now, just as we have done in the above section, we will observe if there are more constraints.  In order to archive  this aim,  we construct the system given in (\ref{eq9}), (\ref{eq10}) and (\ref{eq11}), where $\bar{f}^{}_{AB}$ is given by
\begin{eqnarray*}
\label{eq}
\bar{f}^{}_{AB}=
\left(
  \begin{array}{cccccc}
0   	&\quad 	   0 	  &\quad	0     	 &\quad 	0     	& \quad 	-l\epsilon^{0ab}\delta^{i}{_{j}} 	&\quad	  0  	 \\                                                                        
0	&\quad	   0	   &\quad    0 	  &\quad  	 0 	&\quad 	0   	&\quad  	 0 	\\                                                                   
0 	 &\quad	   0  	 &\quad 	  -2l\epsilon^{0ab}\eta_{ij}     &\quad	   0   &\quad  	 l\epsilon^{0ab}\delta^{i}{_{j}}	 &\quad  		0		\\
0	   &\quad	 	0 	  &\quad 	  0  	  &\quad	0	  &\quad 		0		&\quad 		0 	\\
l\epsilon^{0ab}\delta^{i}{_{j}}	  	&\quad  	0  	&\quad    	-l\epsilon^{0ab}\delta^{i}{_{j}} 	&\quad		  0	 &\quad	0	&\quad 	0 		 \\
0   	&\quad  	 0	  &\quad		0 	 &\quad	0	   &\quad  	0 	&\quad 	 0 	 \\
0   	&\quad  	 0	  &\quad		0 	 &\quad 	0	   &\quad  	l\delta^{i}{_{j}}\epsilon^{0ab}\partial_{a} 	&\quad 	 0 	 \\
0   	&\quad  	 0	  &\quad	-2l\eta_{ij}\epsilon^{0ab}\partial_{b} 	 &\quad 	0	   &\quad 	l\delta^{i}{_{j}}\epsilon^{0ab}\partial_{a} 	&\quad 	 0 	 \\
-l\delta^{i}{_{j}}\epsilon^{0ab}\partial_{b} 		&\quad  	 0	  &\quad	l\epsilon^{0ab}\delta^{i}{_{j}}\partial_{b} 	 &\quad	0	   &\quad  	0 	&\quad	 0 	 \\
 \end{array}
\right)\delta^{2}(x-y).
\end{eqnarray*} \\
This matrix is not a square matrix, but it still has  zero modes. These zero modes are given by
\begin{eqnarray*}
\bar{\mathcal{V}}^{A}_{1} &=& \left(0,\,  V^{A^{i}{_{0}}},\,  0,\,  V^{\tilde{A}^{i}{_{0}}},\,  \partial_{a}V^{i},\,  V^{\lambda_{i0}},\,  0,\,  0,\,  -V^{j} \right), \\
\bar{\mathcal{V}}^{A}_{2} &=& \left(0,\,  V^{A^{i}{_{0}}},\,  - \partial_{a}V^{j},\,  V^{\tilde{A}^{i}{_{0}}},\,  0,\,  V^{\lambda_{i0}},\,  0,\,  V^{j},\,  0 \right), \\
\bar{\mathcal{V}}^{A}_{3} &=& \left(\partial_{a}V^{j},\,  V^{A^{i}{_{0}}},\,  0,\,  V^{\tilde{A}^{i}{_{0}}},\,  0,\,  V^{\lambda_{i0}},\,  V^{j},\, 0,\,  0 \right).
\end{eqnarray*}
By performing the contraction of these zero modes with $Z_{A}$ given by

\begin{eqnarray*}
\label{eq}
Z_{A}=
\left(
  \begin{array}{c}
l\epsilon^{0ab} \partial_{a}\lambda_{i0}	\\                                                                        
l\epsilon^{0ab}\partial_{a}\lambda_{ib}\\                                                                   
-\epsilon^{0ab}l\partial_{a}\lambda_{i0} + 2l\epsilon^{0ab}\partial_{a}\tilde{A}^{i}{_{0}}	\\
-l\{\tilde{F}_{iab}\epsilon^{0ab} + \partial_{a}\lambda_{ib}\epsilon^{0ab}\}	\\
l\partial_{a}A^{i}{_{0}}\epsilon^{0ab} + \partial_{a}\tilde{A}^{i}{_{0}}\epsilon^{0ab} 		 \\
\frac{l}{2}\{F^{i}{_{ab}} - \tilde{F}^{i}{_{ab}}\}\epsilon^{0ab}		\\
0	\\
0	\\
0	\\
 \end{array}
\right)
\end{eqnarray*} \\

we find that the contraction yields identities. Hence,  there are not more constraints. We can observe that in  the FJ formulation there are less constraints than in Dirac's approach (see the Appendix A), however,  if in  the Dirac method we eliminate the  second class constraints, then   the  FJ and Dirac approaches share equivalent  results, the advantage of FJ method   is that it is more economical. The following  step is to add all  previous  information in order to  construct a new symplectic Lagrangian 
\begin{eqnarray}
{\mathcal{L}}^{(1)} &=& l\lambda_{ib}\epsilon^{0ab}\dot{A}^{i}{_{a}} - l \left(\lambda_{ib}\epsilon^{0ab} + \tilde{A}_{ib}\epsilon^{0ab}\right)\dot{\tilde{A}}^{i}{_{a}} - \frac{l}{2}\dot{\Gamma}_{i}\left[F^{i}{_{ab}} - \tilde{F}^{i}{_{ab}}\right]\epsilon^{0ab} - l\dot{\beta}^{i}\partial_{a}\lambda_{ib}\epsilon^{0ab} \nonumber\\ &+& 
  l\dot{\alpha}^{i}\left[\tilde{F}_{iab}\epsilon^{0ab} + \partial_{a}\lambda_{bi}\epsilon^{0ab}\right]- V^{(1)}, 
\label{eq34a}
\end{eqnarray}
where,  $\beta^{i}, \Gamma_{i}$ and $\alpha^{i}$ are Lagrange multipliers enforcing the constraints and the symplectic potential $V^{(1)}=V^{(0)}\mid_{\Omega^{(0)}_{i}=0,\Omega^{(0)}_{2}=0, \Omega^{(0)}_3=0}=0$ vanishes. By following with the method,  from the Lagrangian (\ref{eq34a}) we identify the new set of symplectic variables given by  $\overset{(1)} {\xi}{^A} = (A^{i}{_{a}},\, \beta^{i},\, \tilde{A}^{i}{_{a}},\, \alpha^{i},\, \lambda_{ia},\, \Gamma_{i})$ and the 1-forms $ \overset{(1)}{a}{^{B}} = \left(l\lambda_{ib}\epsilon^{0ab},\, -l\partial_{a}\lambda_{ib}\epsilon^{0ab},\,-l(\lambda_{ib}\epsilon^{0ab} + \tilde{A}_{ib}\epsilon^{0ab}),\,  l[\tilde{F}_{iab}\epsilon^{0ab} + \partial_{a}\lambda_{bi}\epsilon^{0ab}],\, 0,\, -\frac{l}{2}[F^{i}{_{ab}} -  \tilde{F}^{i}{_{ab}}]\epsilon^{0ab}\right)$. By using these symplectic variables, we construct a new symplectic matrix, namely $f^{(1)}_{ab}$, however,    we will find that this symplectic matrix is still singular: again, this suggests  that the theory has a gauge symmetry. In fact,  the gauge symmetry is given by the following Abelian transformations 
\begin{eqnarray}
A_{\mu}^i\rightarrow A^i_{\mu} + \partial_\mu \kappa^i,  \nonumber \\
\tilde{A}_\mu^i\rightarrow \tilde{A}_\mu^i+ \partial_\mu \tilde{\kappa}^i, 
\end{eqnarray}
where $\kappa^i$ and $\tilde{\kappa}^i$ are gauge parameters. Furthermore,  in order to obtain a symplectic tensor, we fix the following gauge
\begin{eqnarray*}
A^{i}{_{0}} &=& 0, \\
\tilde{A}^{i}{_{0}} &=& 0, \\
\lambda_{i0} &=& 0,
\end{eqnarray*}
this implies that $\beta^{i},\, \alpha^{i}$ and $\Gamma_{i}$ are constants.  By introducing the gauge fixing in the Lagrangian, we construct a new symplectic Lagrangian 
\begin{eqnarray}
{\mathcal{L}}^{(2)} &=& l\lambda_{ib}\epsilon^{0ab}\dot{A}^{i}{_{a}} - l\left(\lambda_{ib}\epsilon^{0ab} + \tilde{A}_{ib}\epsilon^{0ab}\right)\dot{\tilde{A}}^{i}{_{a}} - \frac{l}{2}\dot{\Gamma}_{i}\left(F^{i}{_{ab}}\epsilon^{0ab} - \tilde{F}^{i}{_{ab}}\epsilon^{0ab}\right) - \Gamma_{i}\dot{r}^{i} - l\dot{\beta}^{i}\partial_{a}\lambda_{ib}\epsilon^{0ab} \nonumber \\ &-& \beta^{i}\dot{\omega}_{i} + l\dot{\alpha}^{i}\left(\tilde{F}_{iab}\epsilon^{0ab} + \partial_{a}\lambda_{ib}\epsilon^{0ab}\right) - \alpha^{i}\dot{\rho}_{i}.
\end{eqnarray}
where $r^i$, $\omega_i$ and $\rho_i$ are Lagrange multipliers enforcing the gauge conditions. 
Now, we identify the following symplectic variables
$\overset{(2)}\xi{^{A}} =  \left(A^{i}{_{a}},\, \beta^{i},\, \tilde{A}^{i}{_{a}},\, \alpha^{i},\, \lambda_{ia},\, \Gamma_{i},\, r^{i},\, \omega_{i},\, \rho_{i}\right)$  and the one-forms
$\overset{(2)} {a}{^{B}} = \left(l\lambda_{ib}\epsilon^{0ab}, -l\partial_{a}\lambda_{ib}\epsilon^{0ab}, -l[\lambda_{ib}\epsilon^{0ab} + \tilde{A}_{ib}\epsilon^{0ab}], l[\tilde{F}_{iab}\epsilon^{0ab} + \partial_{a}\lambda_{bi}\epsilon^{0ab}], 0,-\frac{l}{2}[F^{i}{_{ab}} - \tilde{F}^{i}{_{ab}}]\epsilon^{0ab},
 -\Gamma_{i}, -\beta^{i}, -\alpha^{i}\right)$.  By using these symplectic variables, we obtain the following symplectic matrix

\begin{eqnarray*}
\label{eq}
f^{(2)}_{AB}=
\left(
  \begin{array}{ccccccccc}
0   	&\,  0 	  &\,	0     	 &\,	0     	&\, 	-l\delta^{i}{_{j}}\epsilon^{0ab} 	&\,  -l\epsilon^{0ab}\delta^{i}{_{j}}\partial_{a} 		&\,		0 	  &\,	  0  	  &\,	0	 \\                                                                        
0	&\,	   0	   &\,   0 	  &\,	 0 	&\,	 l\delta^{i}{_{j}}\epsilon^{0ab}\partial_{a}	   	&\, 	 0	&\,	 0	 &\,	 -\delta^{i}{_{j}} 	&\,  	 0 	\\                                                                   
0 	 &\,	   0  	 &\, 	  -2l\eta_{ij}\epsilon^{0ab}     &\,	   2l\delta^{i}{_{j}}\epsilon^{0ab} \partial_{a}	  &\, 	 -l\delta^{i}{_{j}}\epsilon^{0ab}	 &\, 		l\delta^{i}{_{j}}\epsilon^{0ab} \partial_{a}	&\,	 0	&\,  	 0	&\, 	 0	\\
0	   &\,	 	0 	  &\, 	-2l\eta_{ij}\epsilon^{0ab}\partial_{a}	    &\,	0	  &\, 	-l\delta^{i}{_{j}}\epsilon^{0ab}\partial_{a}		&\, 	0	&\,	0	&\,	0	&\,		-\delta^{i}{_{j}} 	\\
l\delta^{i}{_{j}}\epsilon^{0ab}	  	&\,  	l\delta^{i}{_{j}}\epsilon^{0ab}\partial_{a}	  	&\,   
-l\delta^{i}{_{j}}\epsilon^{0ab} 	&\,	  l\delta^{i}{_{j}}\epsilon^{0ab}\partial_{a}		 &\,	0	&\,	0 	&\, 	0 	&\,	0 	&\,	0 		 \\
l\delta^{i}{_{j}}\epsilon^{0ab}\partial_{a}		&\, 	 0	  &\,	-l\delta^{i}{_{j}}\epsilon^{0ab}\partial_{a} 	 &\,	0	   &\, 	0 	&\, 	 0 	&\,	-\delta^{i}{_{j}}	 &\,  	0	 &\,  	0	 \\
0   	&\, 	 0	  &\,	0 	 &\, 	0	   &\, 	0	&\, 	 \delta^{i}{_{j}} 	&\,	0	&\,	0	&\,	0	 \\
0   	&\,	 \delta^{i}{_{j}}	  &\,	0 	 &\, 	0	   &\, 	0 	&\, 	 0 	&\, 	 0 	&\, 	 0 	&\, 	 0 	 \\
0	&\,	 0	  &\,	0 	 &\,	\delta^{i}{_{j}}	   &\,  	0 	&\,	 0	&\,	 0	&\,		 0	&\,	 0 	 \\
 \end{array}
\right) \nonumber \\
\times \delta^2(x-y), 
\end{eqnarray*} \\
and the inverse is given by
\begin{eqnarray*}
\label{eq}
\left(f^{(2)}_{AB}\right)^{-1}=
\left(
  \begin{array}{ccccccccc}
\frac{1}{2l}\eta^{ij}\epsilon_{ab}  	&\quad  0 	  &\quad	-\frac{1}{2l}\eta^{ij}\epsilon_{ab}      	 &\quad	0     	&\quad 	\frac{1}{l}\eta^{i}{_{j}}\epsilon_{ab} 	&\quad  0 		&\quad	0 	  &\quad	  -\frac{1}{l}\eta^{i}{_{j}}\partial_{a} 	  &\quad	0	 \\                                                                        
0	&\quad	   0	   &\quad   0 	  &\quad	 0 	&\quad		0   	&\quad 	 0	&\quad		 0	 &\quad		 \eta^{i}{_{j}} 	&\quad  	 0 	\\                                                                   
\frac{1}{2l}\eta^{ij}\epsilon_{ab}	 &\quad	   0  	 &\quad 	-\frac{1}{2l}\eta^{ij}\epsilon_{ab}       &\quad	   0	  &\quad 	0	 &\quad	0	&\quad		 0	&\quad  	 0	&\quad 	 -\eta^{i}{_{j}}\partial_{a}		\\
0	   &\quad	 	0 	  &\quad 	0	    &\quad	0	  &\quad 	0	&\quad 	0	&\quad		0	&\quad		0	&\quad		\eta^{i}{_{j}} 	\\
-\frac{1}{l}\eta^{i}{_{j}}\epsilon_{ab}	  	&\quad  0  	&\quad   0	&\quad		0	&\quad		0	&\quad		0 	&\quad		\eta^{i}{_{j}}\partial_{a} 	&\quad		0 	&\quad		0 		 \\
0	&\quad 	 0	  &\quad	0	 &\quad	0	   &\quad 	0 	&\quad		 0 	&\quad	\eta^{i}{_{j}}	 &\quad 	0	 &\quad  	0	 \\
0   	&\quad 	 0	  &\quad	0 	 &\quad 	0	   &\quad 	-\eta^{i}{_{j}}\partial_{a}	&\quad	 -\eta^{i}{_{j}} 	&\quad		0	&\quad		0	&\quad		0	 \\
\frac{1}{l}\eta^{i}{_{j}}\partial_{a}   	&\quad	 -\eta^{i}{_{j}}	  &\quad	0 	 &\quad 	0	   &\quad 	0 	&\quad		 0 	&\quad		 0 	&\quad 	 0 	&\quad 	 0 	 \\
0	&\quad		 0	  &\quad	\delta^{i}{_{j}}\partial_{a} 	 &\quad	-\eta^{i}{_{j}}	   &\quad  	0 	&\quad		 0	&\quad		 0	&\quad		 0	&\quad	 0 	 \\
 \end{array}
\right) \nonumber \\ 
 \times \delta^2(x-y), 
\end{eqnarray*} 

where we can identify the following nonzero FJ  generalized brackets
\begin{eqnarray}
\{ A^{i}{_{a}}(x), A^{j}{_{b}}(y)\}_{FJ} &=& \frac{\eta^{ij}}{2l}\epsilon_{ab}\delta^{2}(x-y), \nonumber \\
\{ A^{i}{_{a}}(x), \tilde{A}^{j}{_{b}}(y)\}_{FJ} &=& -\frac{\eta^{ij}}{2l}\epsilon_{ab}\delta^{2}(x-y), \nonumber \\
\{ A^{i}{_{a}}(x), \lambda_{jb}(y)\}_{FJ} &=& \frac{\delta^{i}_j}{l}\epsilon_{ab}\delta^{2}(x-y), \nonumber  \\
\{ \tilde{A}^{i}{_{a}}(x), \tilde{A}^{j}{_{b}}(y)\}_{FJ} &=& \frac{\eta^{ij}}{2l}\epsilon_{ab}\delta^{2}(x-y).
\label{FJB}
\end{eqnarray}
We can observe that the generalized brackets for TMG and for TMG at the chiral point, in general,  are different,   as expected. On the other hand, in \cite{17} the Dirac brackets between the dynamical variables for TMG  were  not reported, hence, from our results   we  expect that the generalized FJ  brackets  of  TMG will share  a similar  structure  just like that given in (\ref{FJB}).\\
Finally, we  carry out the counting of physical degrees of freedom. There are 18 canonical variables given by $(A^{i}{_{a}}, \tilde{A}^{i}{_{a}},   \lambda_{ia})$  and the following 18 FJ constraints $(\Omega^{(0)}_{1}, \Omega^{(0)}_{2}, \Omega^{(0)}_{3},  A^{i}{_{0}},  \tilde{A}^{i}{_{0}}, \lambda_{i0}  )$, therefore the theory lacks physical degrees of  freedom and it  is topological. \\

\section{ Conclusions and prospects}
In this paper a full FJ approach for an Abelian analog of  TMG and TMG at the chiral point  has been performed. For both theories  
the complete set of FJ  constraints   were found and the quantum brackets identified with the FJ brackets have been constructed. Moreover, we observed that in the  FJ framework   there are less constraints than in the conventional canonical formalism, and  the construction of the FJ brackets   is more economical   and gives the desired Dirac brackets. In addition, we have calculated the number of physical degrees of freedom concluding that the theories  are devoid of physical degrees of freedom and therefore the theories  are topological. The results of this paper are preliminary for performing the quantization  because the principal cornerstone in future works will be the symplectic study of  non-abelian TMG with cosmological constant at the chiral point. However, that work is still in progress and will be the subject of forthcoming works.

\noindent \textbf{Acknowledgements}\\[1ex]
The research of AE, RCF and AHA was supported by Vicerrector{\'i}a de Investigaci\'on y Estudios de Posgrado, BUAP.
AE acknowledges financial support by CONACyT under Grant No. CB-2014-01/ 240781. AHA also thanks PRODEP for partial financial support. All authors thank SNI M\'exico for financial support.
\section{Appendix A}

In this appendix, we will summarize  the Dirac analysis of the Lagrangian  (\ref{eq33}). By performing a full Dirac's analysis, we find the following results; there are 18 first class constraints
\begin{eqnarray}
\tilde{\gamma}^{0}{_{i}} &:& \tilde{\Pi}^{0}{_{i}}\approx 0, \nonumber \\
\gamma^{0}{_{i}} &:& \Pi^{0}{_{i}} \approx 0, \nonumber \\
\bar{\gamma}^{i0} &:& \gamma^{i0} \approx 0, \nonumber \\
 \gamma^{i} &:& \left[F^{i}{_{ab}} - \tilde{F}^{i}{_{ab}}\right]\epsilon^{ab} + \frac{2}{l}\partial_{a}P^{ia} + 2\partial_{a}\chi^{ia} - \frac{1}{2l}\partial_{a}\tilde{\chi}^{ia}  \approx 0 , \nonumber \\
\rho_{i} &:& \frac{1}{2}\tilde{F}_{iab}\epsilon^{ab} - \frac{1}{l}\partial_{a}\tilde{\Pi}^{a}{_{i}} + \frac{1}{l}\partial_{a}P^{a}{_{i}} - \frac{1}{2l}\partial_{a}\chi^{a}{_{i}} - \frac{1}{l}\partial_{a}\tilde{\chi}^{a}{_{i}} \approx 0, \nonumber \\
\beta_{i} &:& \partial_{a}\Pi^{a}{_{i}}\approx 0,
\end{eqnarray}
and the following 12 second class constraints
\begin{eqnarray}
\tilde{\chi}^{a}{_{i}} &:& \tilde{\Pi}^{a}{_{i}} + l\left(\lambda_{ib}\epsilon^{ab} + \tilde{A}_{ib}\epsilon^{ab}\right) \approx 0, \nonumber  \\
\chi^{a}{_{i}} &:& \Pi^{a}{_{i}} - l\lambda_{ib}\epsilon^{ab} \approx 0, \nonumber  \\
\bar{\chi}^{ib} &:& P^{ib}\approx 0,
\end{eqnarray}
where $\left(\Pi^{0}{_{i}}, \tilde{\Pi}^{0}{_{i}}, P^{i0}, P^{ia}, \Pi^{a}{_{i}}, \tilde{\Pi}^{a}{_{i}}\right)$ are the canonical momenta of the dynamical variables $\left(A^{0}{_{i}}, \tilde{A}^{0}{_{i}}, \lambda_{i0}, \lambda_{ia}, A^{i}{_{a}}, \tilde{A}^{i}{_{a}}\right)$ respectively. It is important to comment, that in the Dirac framework there are always  more constraints than in the FJ formalism. However, if we eliminate the second class constraints  by introducing the Dirac brackets, then     equivalent  results are obtained. \\
Hence,  we will construct the Dirac brackets by eliminating only  the second class constraints. In order to perform this aim,  we construct the following matrix whose entries are given by the Poisson brackets between the second class constraints, say, $C_{\lambda \nu}$
\begin{eqnarray*}
\label{eq}
C_{\lambda \nu}=
\left(
  \begin{array}{ccc}
2l\eta_{ij}\epsilon^{ab}		&\quad		0		&\quad		l\epsilon^{ag}\delta^{i}{_{j}} \\
0		&\quad		0		&\quad			-l\epsilon^{ag}\delta^{i}{_{j}} \\
-l\epsilon^{ab}\delta^{i}{_{j}}		&\quad			l\epsilon^{ab}\delta^{i}{_{j}}		&\quad		0 \\
 \end{array}
\right)\delta^{2}(x-y),
\end{eqnarray*}\\
and the inverse is given by
\begin{eqnarray*}
\label{eq}
\left(C_{\alpha \nu}\right)^{-1}=
\left(
  \begin{array}{ccc}
\frac{\eta^{ij}}{2l}\epsilon_{ab}		&\quad		\frac{\eta^{ij}}{2l}\epsilon_{ab}		&\quad		0	\\
-\frac{\eta^{ij}}{2l}\epsilon_{ab}		&\quad		\frac{\eta^{ij}}{2l}\epsilon_{ab}		&\quad		\frac{1}{l}\epsilon_{ab}\delta^{i}{_{j}} \\
0		&\quad		-\frac{1}{l}\epsilon_{ab}\delta^{i}{_{j}}		&\quad		0 \\
 \end{array}
\right)\delta^{2}(x-y).
\end{eqnarray*}\\
In this manner, the Dirac bracket between two functionals, $F$ and $B$, is given by
\begin{eqnarray*}
\{F,B\}_{D} &=& \{F,B\} - \int \{F,\chi_{\alpha}(u)\}C^{-1}_{\alpha\nu}\{\chi_{\nu}(v),B\} du dv,
\end{eqnarray*}
where $\chi_{\alpha}$ represent  the set of second class constraints. Thus, the Dirac brackets between the dynamical variables are given by
\begin{eqnarray}
\{\tilde{A}^{i}{_{a}}(x), \tilde{A}^{j}{_{b}}(y)\}_{D} &=&  \frac{\eta^{ij}}{2l}\varepsilon_{ab}\delta^{2}(x-y), \nonumber \\
\{A^{i}{_{a}}(x), A^{j}{_{b}}(y)\}_{D} &=& \frac{\eta^{ij}}{2l}\epsilon_{ab}\delta^{2}(x-y), \nonumber \\
\{A^{i}{_{a}}, \lambda_{jb}(y)\}_{D} &=& \frac{\delta^{i}{_{j}}}{l}\epsilon_{ab}\delta^{2}(x-y), \nonumber \\
\{ A^{i}{_{a}}(x), \tilde{A}^{j}{_{b}}(y)\}_{D} &=& -\frac{\eta^{ij}}{2l}\epsilon_{ab}\delta^{2}(x-y), \nonumber \\
\{\lambda_{ia}(x), \lambda_{jb}(y)\}_{D} &=& 0,
\end{eqnarray}
where we can see that the Dirac brackets and the FJ brackets given in (\ref{FJB}) coincide to each other.\\
\section{Appendix B}
In this appendix we will resume the FJ approach for singular theories. We start  by writing  the  first order Lagrangian density in the following form \cite{15-c, 15c1, 15c2, 15c3, 15c4, 15c5, 15c6, 15c7}
\begin{equation}
L^{(0)}= a_A \dot{\xi}^{A} - V^{(0)}, 
\end{equation}
here, $a_{A}$ are functions of the  field variables $\xi^{A}$, $V^{(0)}=V^{(0)}(\xi^{A})$ is called the symplectic potential and $A=1, 2, 3...$  labels the number of dynamical variables. Hence, from the Lagrangian density  it is straightforward to show that the  equations of motion  are given by 
\begin{equation}
f^{(0)}_{AB} \dot {\xi}^{B}= \frac{\delta V^{(0)}}{\delta \xi^A}, 
\label{40}
\end{equation}
where 
\begin{equation}
f^{(0)}_{AB} = \frac{\delta a_A }{\delta \xi^{B}} - \frac{\delta a_B }{\delta \xi^{A}}, 
\label{41}
\end{equation}
is the antisymmetric symplectic matrix. Since  the theory is a singular system, there will be constraints.  In this manner, the symplectic matrix (\ref{41}) is not invertible and that means that there are null vectors. In fact,  we call to $\mathcal{V}^{(0)}_A$ the set of null vectors of the symplectic matrix. Hence, from the contraction of the null vectors with the equation (\ref{40}) the following constraints arise  \cite{15-c, 15c1, 15c2, 15c3, 15c4, 15c5, 15c6, 15c7}
\begin{equation}
\Omega^{(0)} = \mathcal{V}^{(0)}_A \frac{\delta V^{(0)}}{\delta \xi^A}=0.
\end{equation}
In analogy with the Dirac method, we demand consistency, this means that the constraints must  be preserved  in time, namely 
\begin{equation}
\dot{\Omega}^{(0)}= \frac{\delta \Omega^{(0)}}{\delta \xi^A } \dot{\xi}^A, 
\label{43}
\end{equation}
hence, by considering (\ref{40}) and (\ref{43}) we can form a  system of linear equations with all information found, say 
\begin{equation}
\left\lbrace
\begin{array}{ll}
 f^{(0)}_{AB} \dot {\xi}^{B} =& \frac{\delta V^{(0)}}{\delta \xi^A},  \\
\frac{\delta \Omega^{(0)}}{\delta \xi^A } \dot{\xi}^A=& 0,
\end{array}
\right.
\label{sys}
\end{equation}
and to rewrite (\ref{sys}) as 
\begin{eqnarray}
\bar{f}_{AB}\dot{\xi}^{B}&=& Z_{A}(\xi), 
\label{45}
\end{eqnarray}
with 
\begin{eqnarray}
\label{eq}
 \bar{f}_{AB}=
\left(
\begin{array}{cc} 
f^{(0)}_{AB} \\ 
\frac{\delta\Omega^{(0)}_{}}{\delta\xi^{A}} 
\end{array}
\right),
\label{46}
\end{eqnarray} 

and

\begin{eqnarray}
\label{eq}
 Z_{A}(\xi)=
\left(
\begin{array}{ccc} 
\frac{\delta V^{(0)}}{ \delta \xi^C} \\ 
0 \\
0\\ 
\end{array}
\right).
\label{47}
\end{eqnarray}
Hence, we repeat the algorithm, we calculate the null vectors of the matrix  (\ref{46}), say $\mathcal{V}^{(1)}$,  and we perform the contraction of these null vectors with $Z_{A}(\xi)$ in order to identify new constraints, 
 \begin{equation}
\Omega^{(1)} = \mathcal{V}{^{(1)}}^A Z_A=0.
\end{equation}
Similarly, by  demanding the  consistency condition 
\begin{equation}
\dot{\Omega}^{(1)}= \frac{\delta \Omega^{(1)}}{\delta \xi^A } \dot{\xi}^A=0, 
\label{49}
\end{equation}
we can combine  (\ref{49}) with equation  (\ref{sys}) to construct a new set of linear equations. By using  these linear equations one verifies   step by step whether there are  new constraints, until there are no more  and we get the identity. \\
On the other hand, we will assume that all  FJ constraints have been identified, hence the final symplectic  Lagrangian can be written as
\begin{equation}
L= a(\xi)_A \dot{\xi}^A - \dot {\gamma}_C \Omega^C - V(\xi), 
\label{50}
\end{equation}
where $\gamma_C$ are Lagrange multipliers enforcing the FJ constraints $\Omega's$ and $V(\xi)= V^{(0)}_{| \Omega^C=0}$. In this manner, if we consider the field variables and the Lagrange multipliers as our new set of symplectic variables, say $\tilde{\xi}^A =(\xi^A, \gamma^B)$, then   we can construct  the  symplectic matrix  of the Lagrangian (\ref{50}) 
\begin{equation}
\tilde{f}_{AB} = \frac{\delta \tilde{a}_A }{\delta \tilde{\xi}^{B}} - \frac{\delta \tilde{a}_B }{\delta \tilde{\xi}^{A}}, 
\label{51}
\end{equation}
where $\tilde{a}_A= a_A+ \gamma^C \frac{\delta \Omega^C}{\delta \xi^A}$ \cite{15-c, 15c1, 15c2, 15c3, 15c4, 15c5, 15c6, 15c7}.  If the   symplectic matrix (\ref{51}) is not  singular, then we can calculate its inverse, namely $\tilde{f}_{AB}^{-1}$, thus we can find all the velocities $ \dot{\xi}^A$ and the problem is finished. On the other hand, if the symplectic matrix  (\ref{51}) is still singular, this means  that the theory has a gauge symmetry,   then in order to obtain a symplectic tensor  it is necessary to fix the gauge  just as in the above sections it was performed. In any case, if $\tilde{f}_{AB}$ is invertible, then we can identify the generalized FJ brackets defining  
\begin{eqnarray}
\{\tilde{\xi}^{A}(x), \tilde{\xi}^{B}(y)\}_{FD}=[\tilde{f}_{AB}(x,y)]^{-1},
\label{52}
\end{eqnarray}
which allow to write  the equations of motion  (\ref{40}) in Hamiltonian form \cite{15-c, 15c1, 15c2, 15c3, 15c4, 15c5, 15c6, 15c7}.

\end{document}